\documentclass{emulateapj}
\usepackage{latexsym}
\usepackage{graphicx}
\usepackage{fancyheadings}
\usepackage{lscape}
\usepackage{natbib}
\usepackage{apjfonts}
\newcommand{\W}{\hphantom{0}}

\begin{document}
\slugcomment{Accepted for publication in Astrophysical Journal Letters}

\title{Phase Variation in the Pulse Profile of SMC X-1}
\shorttitle{PULSE PROFILE OF SMC X-1}

\author{J. Neilsen\altaffilmark{1,2}, R.C. Hickox\altaffilmark{2}, and
 S.D. Vrtilek\altaffilmark{2}}
\altaffiltext{1}{Kenyon College, Department of Physics, Gambier, OH
43022; neilsenj@kenyon.edu}
\altaffiltext{2}{Harvard-Smithsonian Center for Astrophysics, 60 Garden Street, Cambridge, 
MA 02138; rhickox@cfa.harvard.edu; saku@head.cfa.harvard.edu}

\begin{abstract}
We present the results of the timing and spectral analysis of X-ray high-state observations of the high-mass X-ray pulsar SMC X-1 with
\textit{Chandra}, \textit{XMM-Newton}, and \textit{ROSAT}, taken between
1991 and 2001. The source has $L_{\rm X} \sim (3$--$5) \times
10^{38}$ ergs s$^{-1}$, and the spectra can be modeled as a power law
plus blackbody with $kT_{\rm BB}\sim0.18$ keV and a reprocessed emission
radius $R_{\rm BB}\sim 2 \times 10^{8}$ cm, assuming a distance of 60
kpc to the source. Energy-resolved pulse profiles show several
distinct forms, more than half of which include a second pulse in the
soft profile, previously documented only in hard energies. We also
detect significant variation in the phase shift between hard and soft
pulses, as has recently been reported in Her X-1. We suggest an
explanation for the observed characteristics of the soft pulses in
terms of precession of the accretion disk.
\end{abstract}
\section{INTRODUCTION}


SMC X-1 is a high-mass X-ray binary system consisting of an X-ray
source, a $1.6$ $M_{\sun}$ pulsar, accreting from a $17.2$ $M_{\sun}$
companion, Sk 160 \citep{clar03}. It is also the only X-ray pulsar
for which no spin-down episodes have been observed \citep{kaha99}. The
X-ray pulsar has a spin period of $0.71$ s and an eclipsing orbital
period of $3.89$ days, and in addition it shows a superorbital variation in
flux with period between 45 and 60 days. This is likely due to
precession of a warped accretion disk \citep{wojd98}.

The presence of the accretion disk is likely to have significant
consequences for the observed pulse profile. In Her X-1, a low-mass
X-ray pulsar with a regular 35 day superorbital period, reprocessing of
hard X-rays by the disk gives rise to a pulsating, soft spectral
component \citep{endo00}.  \citet{zane04} have shown that the phase
difference between hard and soft pulses changes as the accretion disk
precesses. If similar reprocessing is at work in SMC X-1, as argued by
\citet{hick04}, the pulse profiles should show a variation
similar to that of Her X-1. To that end, we have analyzed pulse
profiles of SMC X-1 in different epochs from \textit{Chandra},
\textit{XMM}, and \textit{ROSAT}. In $\S$ 2 we discuss the
observations and our methods of data analysis. In $\S$ 3 we present
the results of our analysis, and in $\S$ 4 we consider the
implications of our results and future work.

\section{OBSERVATIONS}

\begin{figure}
\plotone{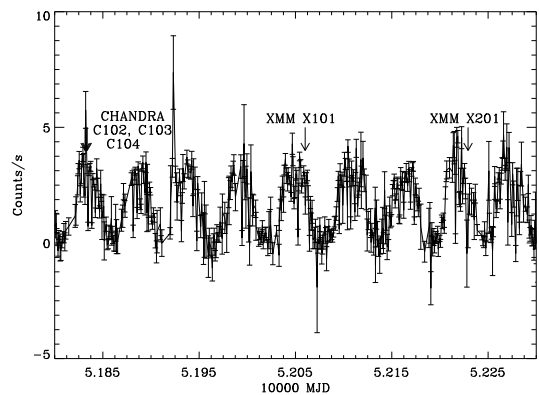}
\caption{\textit{RXTE} ASM lightcurve for SMC X-1 with arrows designating
several observations. Quick-look results as provided by the \textit{RXTE}/ASM team.}
\end{figure}

\begin{deluxetable*}{lccccc}
\tablewidth{4.5in}
\tablecaption{Observations of SMC X-1
\label{tbl-1}}
\tablehead{
\colhead{Observation}  &
\colhead{Observation}  & 
\colhead{Mission}  &
\colhead{Start Time}  & 
\colhead{Observation}  & 
\colhead{Period}  \\
\colhead{ID Number}  &
\colhead{Reference}  &
\colhead{}  &
\colhead{(JD-2400000)}  &
\colhead{Time (ks)} & 
\colhead{(s)}}
\startdata
rp400022n00  &  Rn00  &  \textit{ROSAT}    &  48536.67  &  16.6   & 0.709114(2)   \\
rp400022a02  &  Ra02  &  \textit{ROSAT}    &  49141.50  &  11.8   & 0.708598(5)   \\
400102       &  C102  &  \textit{Chandra}  &  51832.17  &  \W6.2  & 0.70567(2)\W  \\
400103       &  C103  &  \textit{Chandra}  &  51833.34  &  \W6.1  & 0.70567(2)\W  \\
400104       &  C104  &  \textit{Chandra}  &  51834.31  &  \W6.5  & 0.70567(1)\W  \\
0011450101   &  X101  &  \textit{XMM}      &  52060.59  &  46.4   & 0.70542(2)\W  \\
0011450201   &  X201  &  \textit{XMM}      &  52229.65  &  40.0   & 0.70522(3)\W
\enddata
\end{deluxetable*}

The observations reported in this Letter are described in Table 1 and
were made with \textit{Chandra}'s ACIS-S in continuous clocking (CC)
mode (CXC Proposers' Observatory Guide v5.0, 2002), the EPIC-pn
detectors on \textit{XMM-Newton} in full window (X101) and small
window (X201) modes ({XMM-Newton Users' Handbook v2.2, 2004), and the
\textit{ROSAT} PSPC detectors in pointing mode (ROSAT Users' Handbook
1996). We shall refer to the observations as C104, X201, Rn00,
etc. (see Table 1). Fig. 1 locates the \textit{Chandra} and
\textit{XMM} observations in the All-Sky Monitor (ASM) lightcurve from
the \textit{Rossi X-ray Timing Explorer} (\textit{RXTE}) mission. We
estimate the superorbital phase ($\phi_{\rm SO}$) as 0.16 for the
\textit{Chandra} observations, 0.37 for X101, and 0.42 for X201, where
$\phi_{\rm SO}=0$ at the beginning of the X-ray high state. Given the
variation in $\phi_{\rm SO}$, the uncertainty for these estimates is
plus or minus a few days or $\phi_{\rm SO} \sim 0.05$. The $\phi_{\rm
SO}$ estimates for \textit{ROSAT} are unavailable because those
observations were taken before the launch of \textit{RXTE}.

Since the CC mode provides only one dimension of spatial information, ACIS
source events were extracted from a source-centered rectangle of size
$0\farcs492 \times 1\farcs968$. Background events were
extracted from two regions $0\farcs492 \times 9\farcs84$
equidistant from the source. For X101, which
shows the source entering an eclipse, events were located within a
circular region of radius 45$\arcsec$ and extracted from the 4.9 ks of
pre-eclipse data; for X201, we used a circular region of radius
$51\farcs2$. We ignored the \textit{XMM} background because its
surface brightness was less than 0.1\% of that of the source
region. \textit{ROSAT} source events were extracted from a circular
region of radius $2\farcm5$. For Ra02, which shows an eclipse,
events were taken from the 5.5 ks of post-eclipse data.

\section{ANALYSIS}

Before the spectroscopic and timing analysis, we performed a barycentric
correction and corrected for the orbital motion of SMC X-1 using the
orbital ephemeris calculated by \citet{wojd98}. The following
analysis was performed with CIAO version 3.1 for \textit{Chandra} and
\textit{ROSAT} data and SAS version 5.3.3 for \textit{XMM}.

\subsection{Phase-Averaged Spectroscopy}

For \textit{Chandra} and \textit{XMM}, we extracted phase-averaged
spectra, modeling the 0.6--9.8 keV emissions with neutral absorption,
a blackbody with $kT_{\rm BB} \sim 0.18$ keV, and a power law.  For the
\textit{XMM} spectra we also included a high-energy cutoff \citep[$E_{\rm
cut} \sim 6$ \rm {keV}, $E_{\rm fold}\sim 8$ \rm {keV}][]{woo95}.  For \textit{Chandra} and X201, we excluded the energy
range 1.7--2.8 keV to avoid instrumental effects (near the Au and Si
edges) for high count data \citep{mill02}. Although the excellent time
resolution of CC mode prevents some photon pileup, we used the ISIS
pileup kernel for the \textit{Chandra} observations \citep{davi01},
which show a pileup fraction of $\sim$17\%. For \textit{XMM} we minimized
pileup in the spectra by excluding events from the central one-third radius
of the extraction regions.

\begin{figure}
\epsscale{1.0}
\plotone{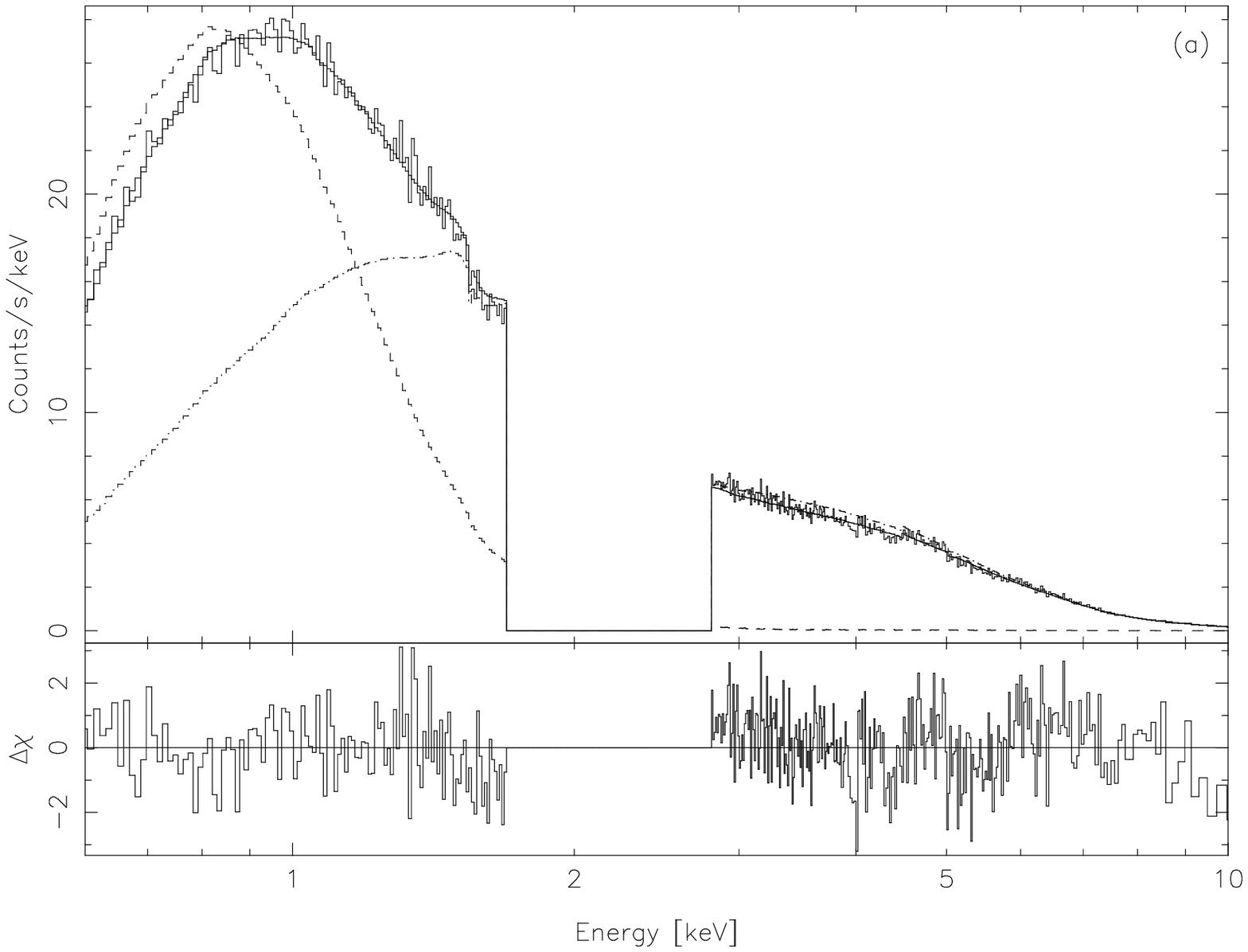}
\plotone{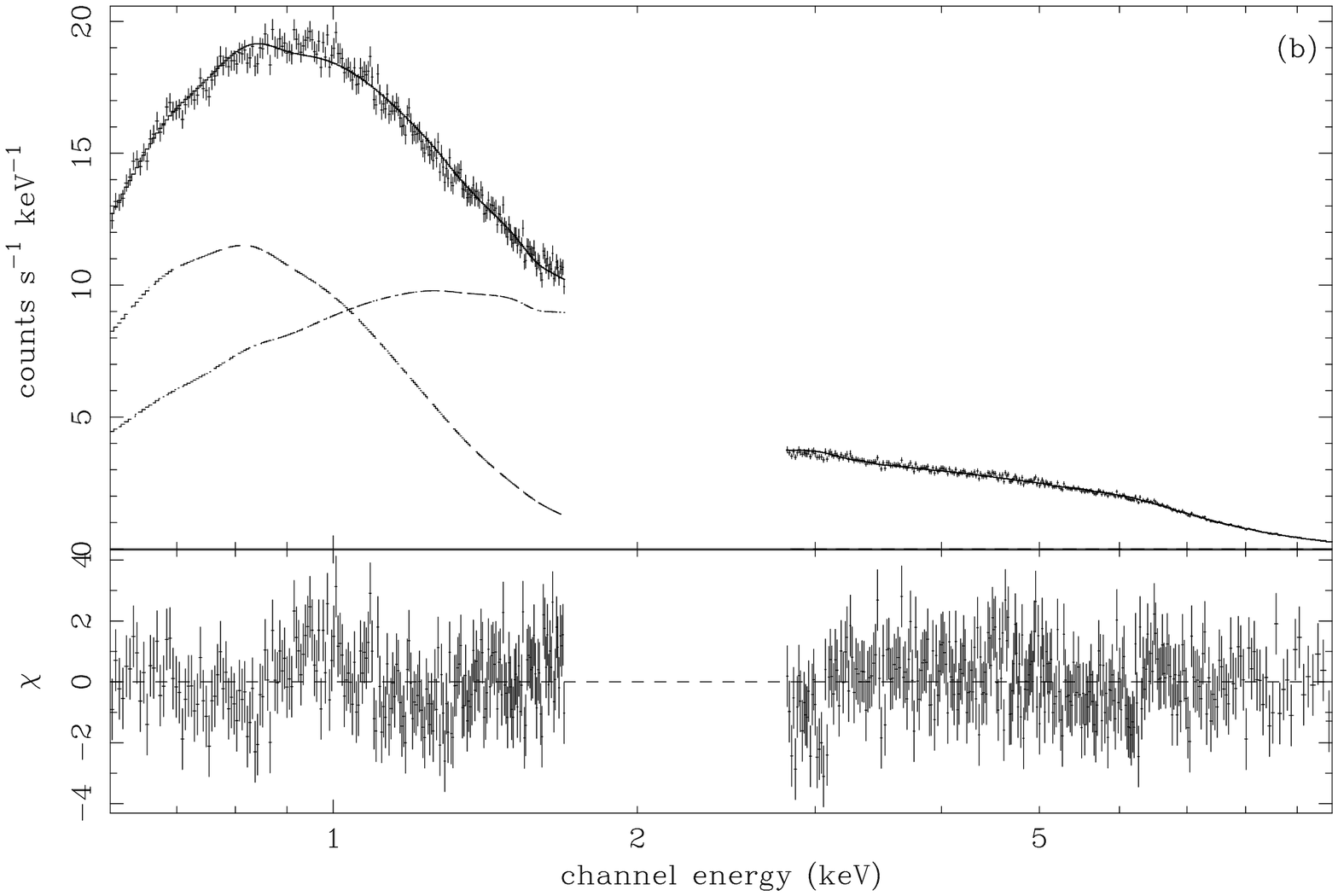}
\plotone{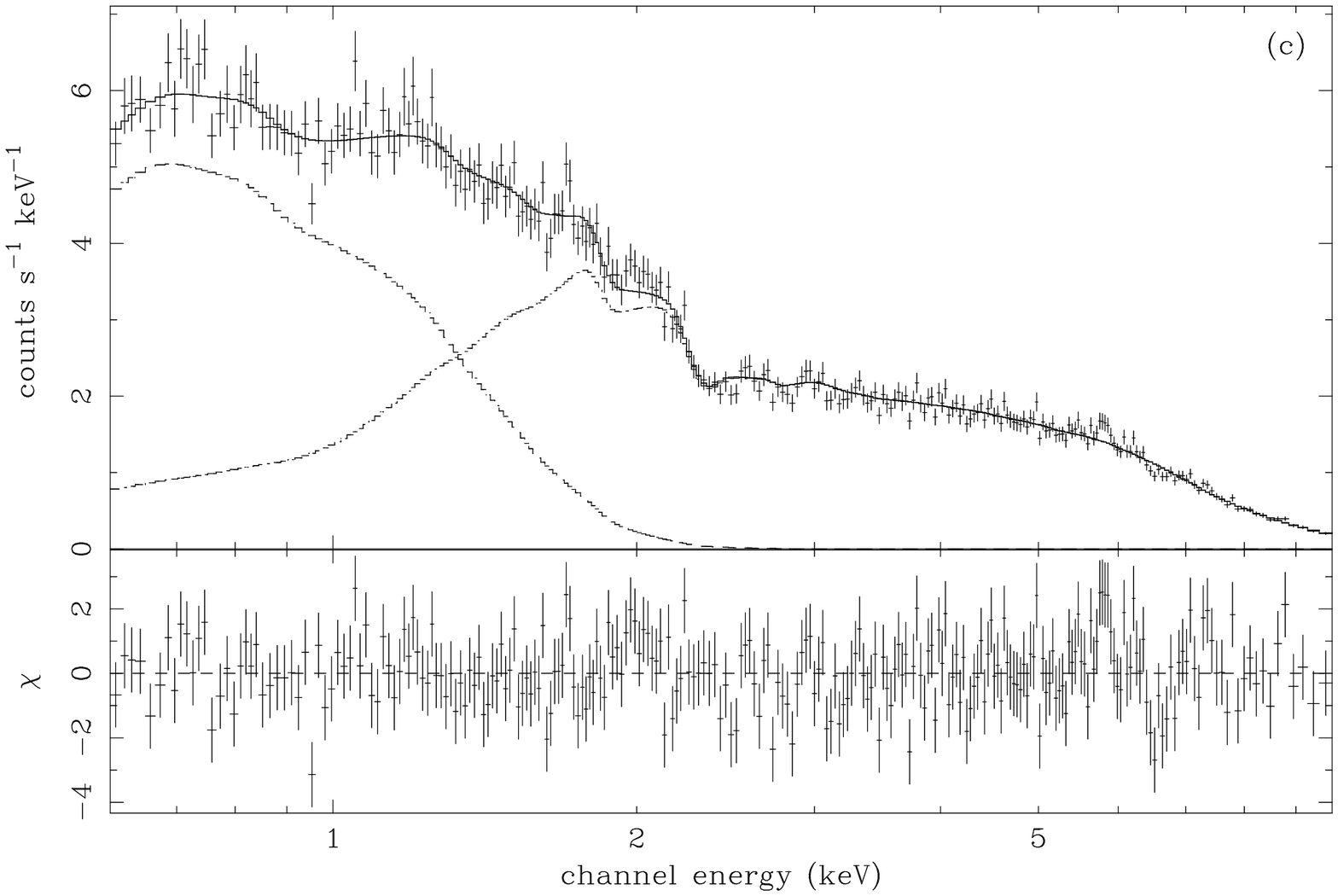}
\caption{Spectra from (a) C104, (b) X201, and (c) X101 with blackbody
and power law components shown.  As X101 is a pre-eclipse observation,
it was fit with an additional partial covering absorption. Note that
in (a) the model components do not add exactly to the observed counts.
This is due to the pileup kernel used in the fit.}
\end{figure}

\begin{figure}
\epsscale{0.95}
\plotone{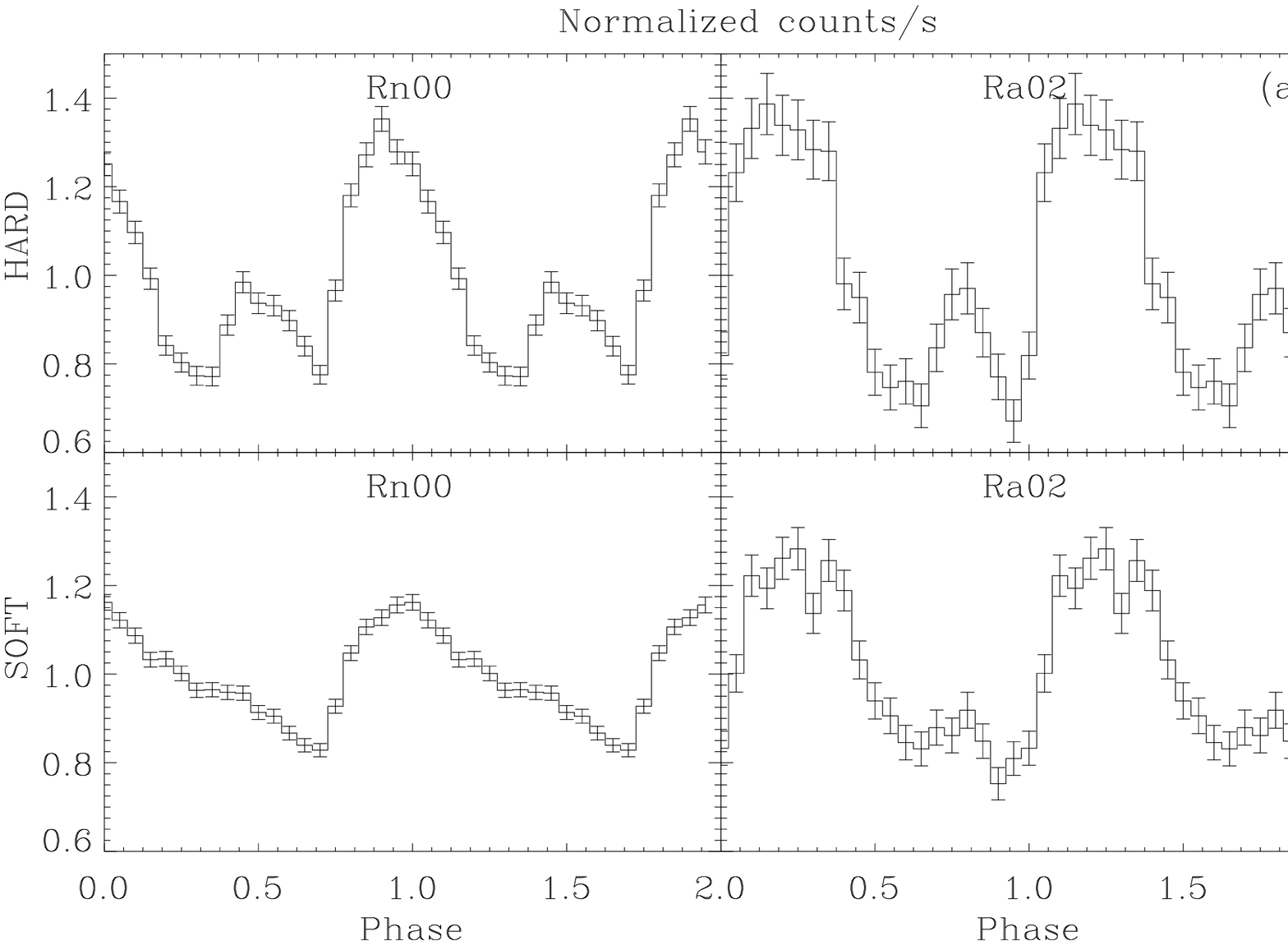}
\plotone{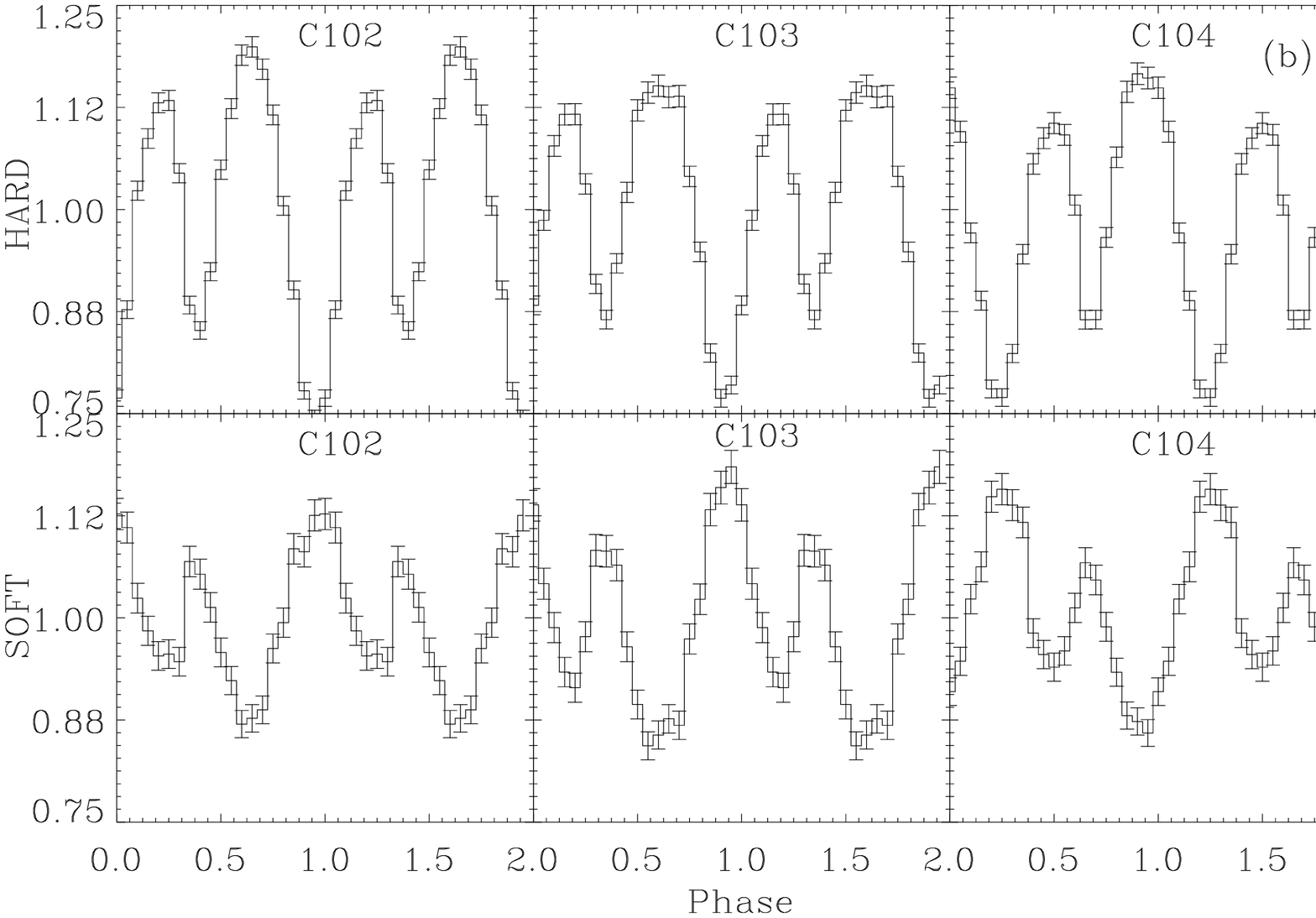}
\plotone{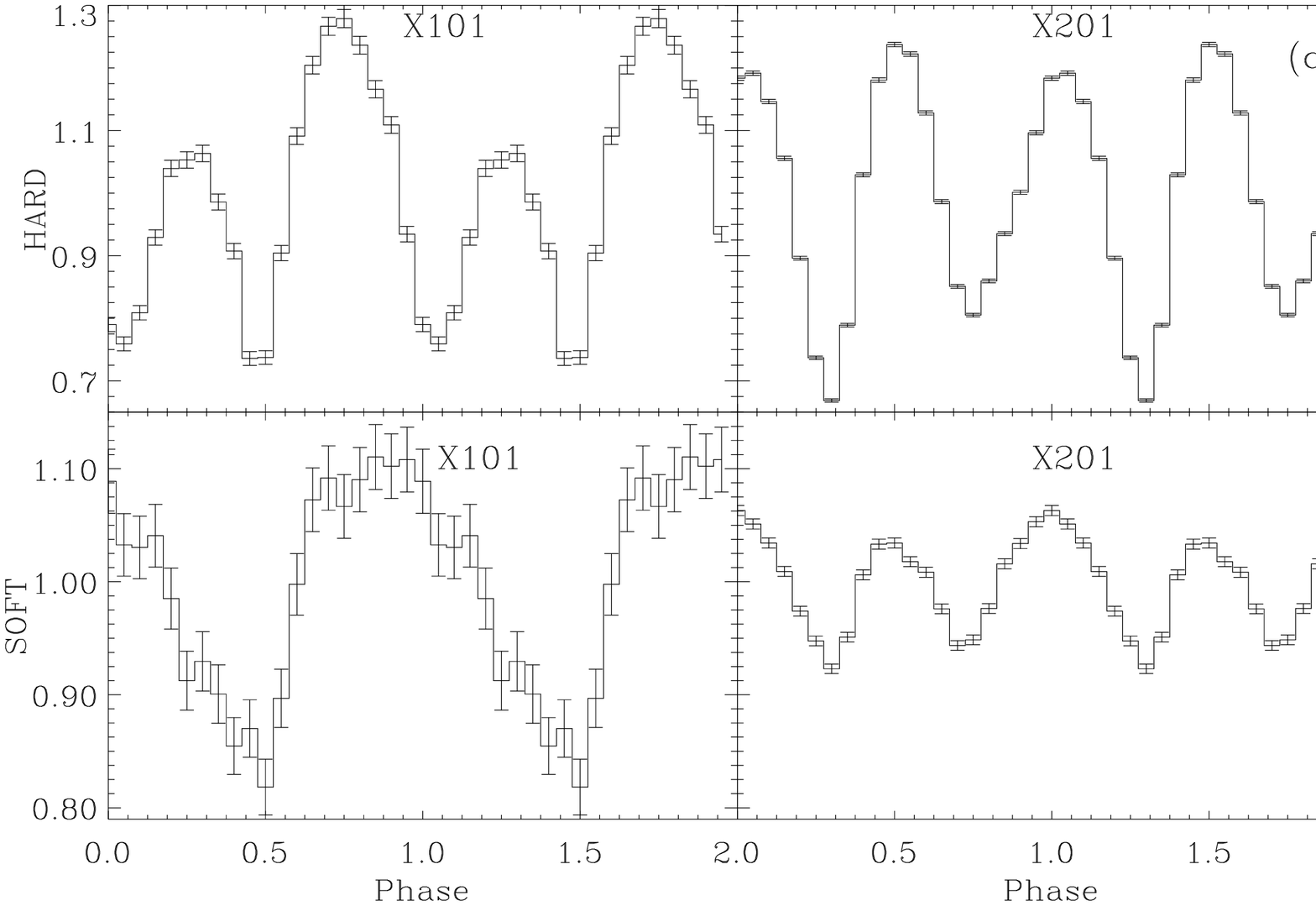}
\caption{Hard and soft pulse profiles for (a) \textit{ROSAT}, (b)
\textit{Chandra}, and (c) \textit{XMM}. The hard range is 2.0--8.0 keV
(1.5--2.4 keV for \textit{ROSAT}); soft is 0.5--1.0 keV. Two pulse
periods are shown for clarity.  Note that the pulse phases shown
are not absolute, but are only relative to each observation.}
\end{figure}

Spectral fits from C104, X201, and X101 are shown in Fig. 2 and
results are listed in Table 2. All spectra are dominated by blackbody
emission below $\sim$1.0 keV, athough the similar wavy residuals in the
\textit{Chandra} and X201 spectra below 1.0 keV suggest that the
soft emission is not exactly blackbody. For X101, the data are taken
as the source is entering eclipse, and the spectrum could not be fitted
using the simple model above.  Fixing $\Gamma$ to a typical value of
0.9, we found that a good fit is achieved by including a partial
covering absorption, with a covering fraction of 88\%. This accounts
for the extra absorption by the dense gas around the star as the eclipse
begins, but allows for some X-rays to be scattered around the
absorbing region. Residuals suggest that there may also be Fe absorption at
$\sim$$6.5$ keV, but more detailed spectral analysis is beyond the
scope of this Letter. Since blackbody emission dominates below 1.0 keV,
and power-law emission above 2.0 keV, we can use energy-resolved profiles in
these ranges to examine the variation of the separate components.
More detailed analysis, using pulse-phase spectroscopy, will be
presented in a forthcoming paper.

\begin{deluxetable*}{lccccc}
\tablecaption{Fits to SMC X-1 Spectra
\label{tbl-2}}
\tablewidth{5.75in}
 \tablehead{ \colhead{Parameter} & \colhead{C102} &
\colhead{C103} & \colhead{C104} & \colhead{X201} & \colhead{X101}}
\startdata 
$N_{\rm H}$  & $0.14^{+0.02}_{-0.02}$ & $0.23^{+0.02}_{-0.02}$ & $0.19^{+0.01}_{-0.01}$ & $0.17^{+0.01}_{-0.01}$ & $0.08^{+0.02}_{-0.02}$ \\
$\Gamma$  & $0.89^{+0.03}_{-0.04}$ & $1.03^{+0.03}_{-0.04}$ & $0.92^{+0.03}_{-0.04}$ & $0.91^{+0.02}_{-0.02}$ & 0.9 \\
$kT_{BB}$  & $0.177^{+0.007}_{-0.007}$ & $0.169^{+0.005}_{-0.005}$ & $0.178^{+0.005}_{-0.005}$ & $0.190^{+0.004}_{-0.004}$ & $0.173^{+0.013}_{-0.010}$ \\
$E_{\rm cut}$ (keV)  & \nodata & \nodata & \nodata & $6.1^{+0.1}_{-0.1}$ & $5.9^{+0.2}_{-0.2}$ \\
$E_{\rm fold}$ (keV)  & \nodata & \nodata & \nodata & $6.8^{+0.3}_{-0.3}$ & $8.9^{+1.1}_{-1.2}$ \\
Pileup fraction  & 0.16 & 0.18 & 0.16 & \nodata & \nodata \\
Partial $N_{\rm H}$ (10$^{22}$cm$^{-2}$) & \nodata & \nodata & \nodata & \nodata & $1.2^{+0.1}_{-0.1}$ \\
Covering fraction & \nodata & \nodata & \nodata & \nodata & $0.88^{+0.02}_{-0.03}$ \\
$\chi^2_{\nu}$ (d.o.f.)  & 1.36 (384) & 1.19 (387) & 1.22 (390) & 1.20 (516)
& 1.20 (258) \\
Observed flux ($10^{-9}$ ergs cm$^{-2}$ s$^{-1}$) & 1.1 & 1.1 & 1.1 & 1.0 & 0.57 \\
Unabsorbed flux ($10^{-9}$ ergs cm$^{-2}$ s$^{-1}$) & 1.2 & 1.2 & 1.3 & 1.1 & 0.84 \\
Luminosity (ergs s$^{-1}$)  & 5.2 & 5.3 & 5.4 & 4.7 & 3.6

\enddata \tablecomments{Errors
are 90\% confidence for a single parameter. Fluxes are for
0.6--9.8 keV.  A distance of 60 kpc is assumed for the SMC.}
\end{deluxetable*}

\subsection{Timing Analysis}

Since SMC X-1 is a fast-period pulsar, pulse profile analysis requires
a time resolution of better than 0.71 s. Fortunately, the \textit{Chandra},
X201, X101, and \textit{ROSAT} observations have sufficient time
resolutions of 2.85, 6, 73.4, and 130 ms, respectively. We used
\textit{efsearch} in the software package XRONOS to measure the pulse
periods (see Table 1).  These are consistent with the spin-up rate
calculated by \citet{wojd98} and the periods determined by
\citet{vrti01} from the same \textit{Chandra} data. We extracted high-
and low-energy pulse profiles from each of the seven observations (see
Fig. 3).  Here we do not correct for pileup effects, since these have minimal impact on the shapes of the pulses.

All seven observations show the double-peaked hard pulse profile that
has been thoroughly documented \citep{wojd98, paul02, naik04}.
The soft profile, however, varies markedly.  While earlier studies of
SMC X-1 have shown a roughly sinusoidal shape of soft pulses
\citep{paul02, naik04}, the \textit{ROSAT}, \textit{Chandra}, and
\textit{XMM} observations show a complex variation of profiles. Although
the \textit{ROSAT} pulse profiles show a single main soft peak, four
of the five \textit{Chandra} and \textit{XMM} observations show
\textit{double}-peaked soft pulses.

The first \textit{ROSAT} observation, Rn00, shows a single asymmetric
hump at lower energies, slightly out of phase with the hard
pulses. For Ra02, the soft pulses are lightly double-peaked, although
the poor spectral resolution of the PSPC places the exact details of the
profiles in doubt.  The X101 pulse profile shows a single broad peak quite
similar to Rn00.  In the \textit{Chandra} data, we
see a clear second peak in all three soft pulse profiles, and we note a
\textit{significant phase difference} between hard and soft pulses. In
X201 we also find double-peaked soft pulses, but these are
\textit{almost in phase} with the hard pulses.

For \textit{Chandra} and \textit{XMM} we cross-correlated the hard and
soft profiles, to quantify the phase shift between the pulses (see
Fig. 4), as in \citet{rams02} and \citet{zane04}. The \textit{Chandra}
profiles show strong anticorrelations at a phase shift of
0$\degr$ and positive correlations at $\pm$90$\degr$. We see
multiple peaks because the hard and soft profiles each have two
pulses.  Observation X101 shows a broad correlation with a peak at
$\sim$20$\degr$; in X201 we detect possible phase shifts of
$\sim$$0\degr$, $\sim\pm90\degr$, and $\sim\pm170\degr$.

\section{DISCUSSION}

\begin{figure}
\epsscale{1.0}
\plotone{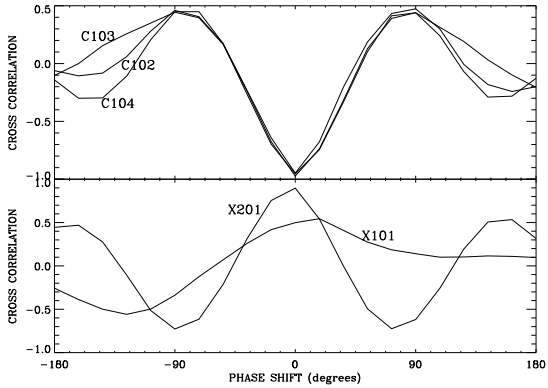}
\caption{Cross-correlation between hard and soft pulse profiles for
\textit{Chandra} and \textit{XMM} data.}
\end{figure}

In these observations both hard and soft X-rays show
pulsations, but differences in the hard and soft pulse profiles
indicate a different geometric or physical origin of emission
\citep{paul02}.  Soft pulses in the most luminous X-ray pulsars are likely to
originate at the inner edge of the accretion disk, where hard pulses
from the neutron star are reradiated at lower energies by disk gases
\citep{hick04}. If this picture is valid for SMC X-1, the observed
blackbody component should be consistent with emission from the inner
disk. For reprocessing by a partial spherical surface around the
neutron star, the blackbody radius $R_{\rm BB}$ is given by
\citep{paul02}
$$R_{\rm BB}=\sqrt{\frac{L_{\rm X}}{4\pi\sigma T^{4}_{\rm BB}}}.$$ Using
$kT_{\rm BB}\sim 0.18$ keV and $L_{\rm X}\sim 4 \times 10^{38}$ ergs
s$^{-1}$, we have $R_{\rm BB}\sim 1.7 \times$10$^{8}$ cm. 

In the standard model for X-ray pulsars, the inner disk radius $R_0$
is close to the magnetospheric radius $R_{m}$, where the magnetic
pressure of the dipole field equals the ram pressure of the infalling
gas.  However, $R_{m}$ is difficult to estimate for SMC X-1,
because the surface $B$-field has not been directly measured. We note
that $R_{\rm BB}$ is close to the corotation radius $R_{\rm cor} =
(GMP^2/4 \pi^2)^{1/3} \simeq 1.3\times 10^{8}$ cm, which is another
estimate for $R_0$.  However, the lack of spin-down episodes for SMC
X-1 suggests that $R_0 < R_{\rm cor}$ \citep{kaha99}. The situation is
therefore unclear, but without additional constraints it is certainly
possible that the soft pulses originate at the inner accretion disk.

In light of this picture we consider the following pulse profile
characteristics:  
\begin{enumerate}
\item The hard pulse profile is consistently double-peaked, typically
with one dominant peak.
\item The soft pulse profile varies in shape,
and has been observed with either a single asymmetric peak or two
peaks with varying degrees of symmetry.
\item The pulse profiles exhibit large-scale variations in phase
relationship. Closely-spaced observations have revealed identical
phase shifts.
\end{enumerate}

\citet{zane04} suggest that the varying hard-soft phase shift in Her
X-1 may be due to the precession of the accretion disk, so we consider
the possibility that such a disk could reproduce the above
characteristics in SMC X-1. If the observational line of sight is
sufficiently close to the plane of the magnetic axis, we might always
see two hard pulses, one from each magnetic pole. However, the soft
profile could change dramatically in shape and relative phase, as
observed, if
precession of the disk causes the visible part of the reprocessing
region to vary. 

We note however that we cannot conclusively verify this picture and
its analogy to Her X-1.  First, we lack sufficient sampling of the
full 60 day superorbital period.  Second, the hard-soft phase shift
appears to vary more rapidly for SMC X-1 (a change of
$\simeq$90$\degr$ between $\phi_{\rm SO} \simeq 0.16$ and 0.42) than
for Her X-1 \citep[$\simeq$20$\degr$ between $\phi_{\rm SO} \simeq
0.03$ and 0.17][] {zane04}.  Third, it is not completely clear even in
Her X-1 that the phase variation is caused by the 35 day disk
precession; soon after the observations of \citet{zane04}, the source
entered an anomalous low state, an event that may be caused by
changes in the inner disk structure \citep{boyd04,oost01}.  Such
changes could alter the shape of the inner reprocessing region, and
thus the soft profiles, in a way that is not directly related to the
superorbital period.

We conclude that in SMC X-1, we have observed variations in the soft
pulse profile that likely reflect changes in the reprocessing of hard
X-rays by the accreting gas.  These variations may be related to the
superorbital precession of the accretion disk, as has been proposed
for Her X-1.  If it can be shown with more observations that the pulse
profiles and superorbital period are correlated, then we will have the
opportunity to apply significant constraints to the geometry of SMC
X-1.

\acknowledgements We thank Cara Rakowski, John Houck, Craig Heinke,
Jonathan McDowell, and Patrick Wojdowski for helpful discussions and
the referee for useful comments. This work was supported in part by
NSF grants 9731923 and AST 0307433 and NASA grant NAG5-10780.

\end{document}